\documentclass[aps, prx, twocolumn, showpacs, superscriptaddress]{revtex4-1}
\usepackage{graphicx}
\usepackage{amsmath}
\usepackage{amssymb}
\usepackage{physics}
\usepackage{hyperref}
\usepackage{lipsum}
\hypersetup{colorlinks=true,
	    final=true,
	    linkcolor=blue,
	    citecolor=blue,
	    filecolor=blue,
	    urlcolor=blue,
}

\begin{document}
\title{Correlated electronic structure of a quintuple-layer nickelate}
\author{Harrison LaBollita}
\affiliation{Department of Physics, Arizona State University, Tempe, AZ 85287, USA}
\author{Antia S. Botana}
\affiliation{Department of Physics, Arizona State University, Tempe, AZ 85287, USA}
\date{\today}

\begin{abstract}
We present a comparative density-functional theory plus dynamical mean-field theory (DFT+DMFT) study of the two known superconducting members of the rare-earth (R) layered nickelate family: hole-doped RNiO$_{2}$ ($n=\infty$) and R$_{6}$Ni$_{5}$O$_{12}$ ($n=5$). At the same nominal carrier concentration, these two materials exhibit nearly identical electronic structures and many-body correlations effects: mass enhancements, self-energies, and occupations. However, the fermiology of the quintuple-layer nickelate is more two-dimensional-like than its infinite-layer counterpart making this new superconducting quintuple-layer nickelate more cuprate-like without the need for chemical doping.
\end{abstract}

\maketitle

%%%%%%%%%%%%%%%%%%% INTRODUCTION %%%%%%%%%%%%%%%%%%%%%%%%%%%%%%%%%%%
\section{\label{sec:intro}Introduction}
Understanding the mechanism behind high-temperature superconductivity (HTS) has been a long-standing challenge since the discovery of cuprates in 1986 \cite{Bednorz1986}. The study of materials with similar layered structures and $3d$ electron count has been perceived as one strategy to help tackle this problem. In this context, nickel oxide materials have been an obvious target for decades given the proximity of Ni and Cu in the periodic table (Ni$^{1+}$ being isoelectronic with Cu$^{2+}$) \cite{Anisimov1999, pickett2004}. 

\begin{figure}[ht]
	\centering
	\includegraphics[width=0.8\columnwidth]{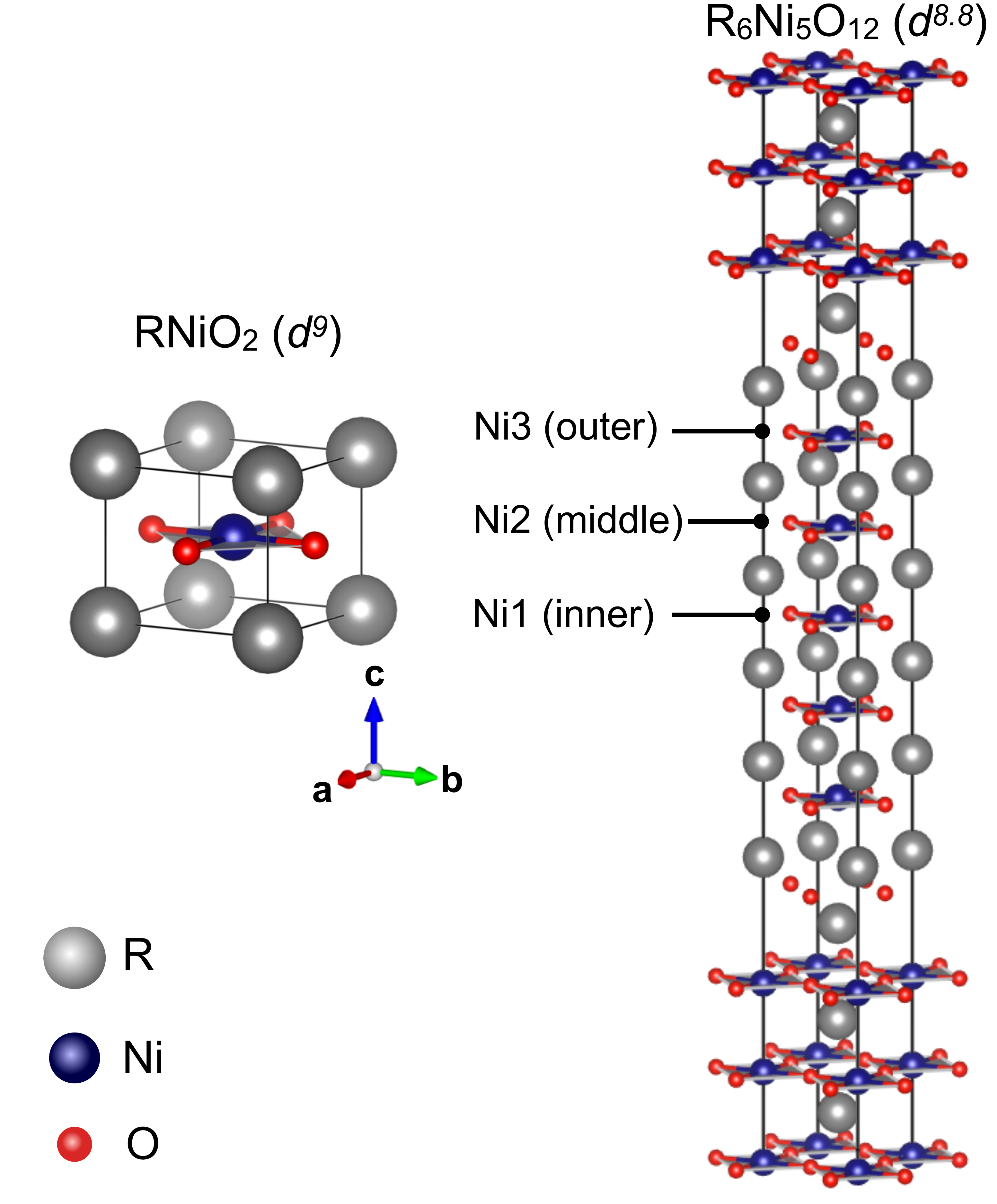}
	\caption{Crystal structure of R$_{n+1}$Ni$_{n}$O$_{2n+2}$ nickelates for $n=\infty$ (left) with space group P4/mmm and $n=5$ (right) with space group I4/mmm. For the 5-layer compound, we highlight the inner, middle, and outer NiO$_{2}$ planes. The separating slab between the five NiO$_{2}$ layers in the $n=5$ material is referred to as a fluorite  blocking layer. Grey, blue, and red spheres denote the R (La), Ni, and O atoms, respectively.}
	\label{fig:struct}
\end{figure}

After a 30-year quest, superconductivity upon hole-doping has been found in hole-doped RNiO$_2$ materials (R= La, Pr, Nd) \cite{Li2019, Osada2020, Osada2021nickelate, Zeng2021superconductivity}, attracting a great deal of experimental  \cite{Li2019, Fu2020, Osada2020, lee2020aspects, goodge2020, hepting2020, Fu2020, Li2020, Li2020dome, BiXiaWang2020, QiangqiangGu2020, cui2020nmr, Liu2020, Osada2021nickelate, Zeng2021superconductivity} and theoretical \cite{wu2019, Hu2019, jiang2019, nomura2019,  Choi2020, Ryee2020, Gu2020, Karp2020_112, botana2020, Leonov2020, Kapeghian2020, lechermann2020late, lechermann2020multi, pickett2020, zhang2020self, Sakakibara2020, jiang2020, werner2020, Zhang2020eff, Zhang2020, Wang2020, Bandyopadhyay2020} attention. These systems have a nominal $d^9$ filling in their parent phase and their structure displays infinite NiO$_2$ planes, in analogy to the CuO$_2$ planes of the cuprates (see Fig. \ref{fig:struct}). Upon hole-doping, superconductivity has been observed in the infinite-layer nickelate with a maximum T$_{c}\sim$ 15 K near $d^{8.8}$ nominal filling, coincidental with optimal doping in the cuprates. Importantly, RNiO$_2$ materials are the infinite-layer ($n=\infty$) members of a larger series of layered nickel oxide compounds, represented by the general chemical formula R$_{n+1}$Ni$_n$O$_{2n+2}$, where $n$ is the number of NiO$_2$ planes along the $c$-axis. Recently, the five-layer ($n=5$) member of the series Nd$_6$Ni$_5$O$_{12}$, also with an average $d^{8.8}$ nominal filling, has been found to be superconducting with a similar T$_c$ but without the need for chemical doping \cite{pan2021super}, a discovery that has opened up the door to a potential whole new family of nickelate superconductors beyond the infinite-layer material. 
 
Here, we present a comparative density-functional theory plus dynamical mean-field theory (DFT+DMFT) study of the correlated electronic structure of the $n=\infty$ and $n=5$ nickelates. We focus on comparing the materials at the same $d^{8.8}$ nominal filling -where superconductivity arises- but also present results for the parent infinite-layer material at $d^9$ nominal filling as a benchmark. Overall, the quintuple-layer and infinite-layer nickelates exhibit similar electronic structures and many of the same correlated features (i.e. electronic self-energies, mass enhancements, and occupations). However, the 5-layer material presents a much more two-dimensional-like electronic structure making it more cuprate-like relative to its infinite-layer counterpart without the need for chemical doping.

%%%%%%%%%%%%%%%%%%% FIG crystal structures%%%%%%%%%%%%%%%%%%%%%%%%%%%%%%%%%%%

%%%%%%%%%%%%%%%%%%% CRYSTAL STRUCTURES %%%%%%%%%%%%%%%%%%%%%%%%%%%%%%%%%%%
\section{\label{sec:crystal}Crystal structures}
All layered-nickelates in the R$_{n+1}$Ni$_n$O$_{2n+2}$ family contain infinite NiO$_2$ planes (see Fig. \ref{fig:struct}) and are derived from a parent perovskite ($n=\infty$) or Ruddlesden-Popper ($n \neq \infty$) phase via oxygen reduction. In the $n= \infty$ material each NiO$_2$ plane is separated by a layer of R ions along the $c$ axis. In the $n= 5$ material, there are five NiO$_2$ planes with the two outer and middle layers being equivalent by symmetry while the inner layer acts as a mirror plane (see Fig. \ref{fig:struct}). Each of these planes is also separated by a layer of R ions but, in addition to the five R-NiO$_{2}$ structural units, the quintuple-layer material has a fluorite blocking R$_2$O$_2$ slab (common to all $n \neq \infty$ materials). Further, each neighboring five layer group is displaced by half a lattice constant along the $x$ and $y$ directions. These two additional structural features effectively decouple the neighboring 5-layer blocks and cut off the $c$-axis dispersion in the 5-layer material with respect to the infinite-layer system \cite{pardo2010}.

%%%%%%%%%%%%%%%%%%% THEORETICAL FRAMEWORK %%%%%%%%%%%%%%%%%%%%%%%%%%%%%%%%%%%
\section{\label{sec:theory}Methods}
Density-functional theory (DFT) calculations are performed using the all electron, full potential code {\sc wien2k} \cite{w2k} based on the augmented plane wave plus local orbital (APW+lo) basis set with the Perdew-Burke-Ernzerhof (PBE) \cite{pbe} implementation of the generalized gradient approximation (GGA) for the exchange-correlation functional. We have chosen to study the two layered nickelates with R $=$ La to avoid ambiguities in the treatment of the $4f$ states that would arise from Nd or Pr. We construct the structure of the La-based 5-layer nickelate using the structure of the La$_4$Ni$_3$O$_8$ material as a reference (tetragonal with an I4/mmm space group) \cite{poltavets2007}. We subsequently optimize the lattice parameters and internal coordinates for each phase within GGA. The in-plane lattice parameters are almost identical for both compounds ($\sim$ 3.97 \AA{}), while the out-of-plane lattice parameter obviously increases with the number of layers ($c=3.37$ \AA{} and $c=39.93$ \AA{} for the infinite-layer and 5-layer materials, respectively). We note that the same structural optimization procedure applied to the Nd-based 5-layer nickelate gives rise to structural parameters that are in excellent agreement with experimental data (see Ref. \onlinecite{pan2021super}). In order to hole-dope  the infinite-layer material to achieve a $d^{8.8}$ nominal filling, we employ the virtual crystal approximation (VCA) applied to the La atoms. 

%%%%%%%%%%%%%%%%%%% FIGURE DFT BANDS %%%%%%%%%%%%%%%%%%%%%%%%%%%%%%%%%%%
\begin{figure}
\centering
\includegraphics[width=\columnwidth]{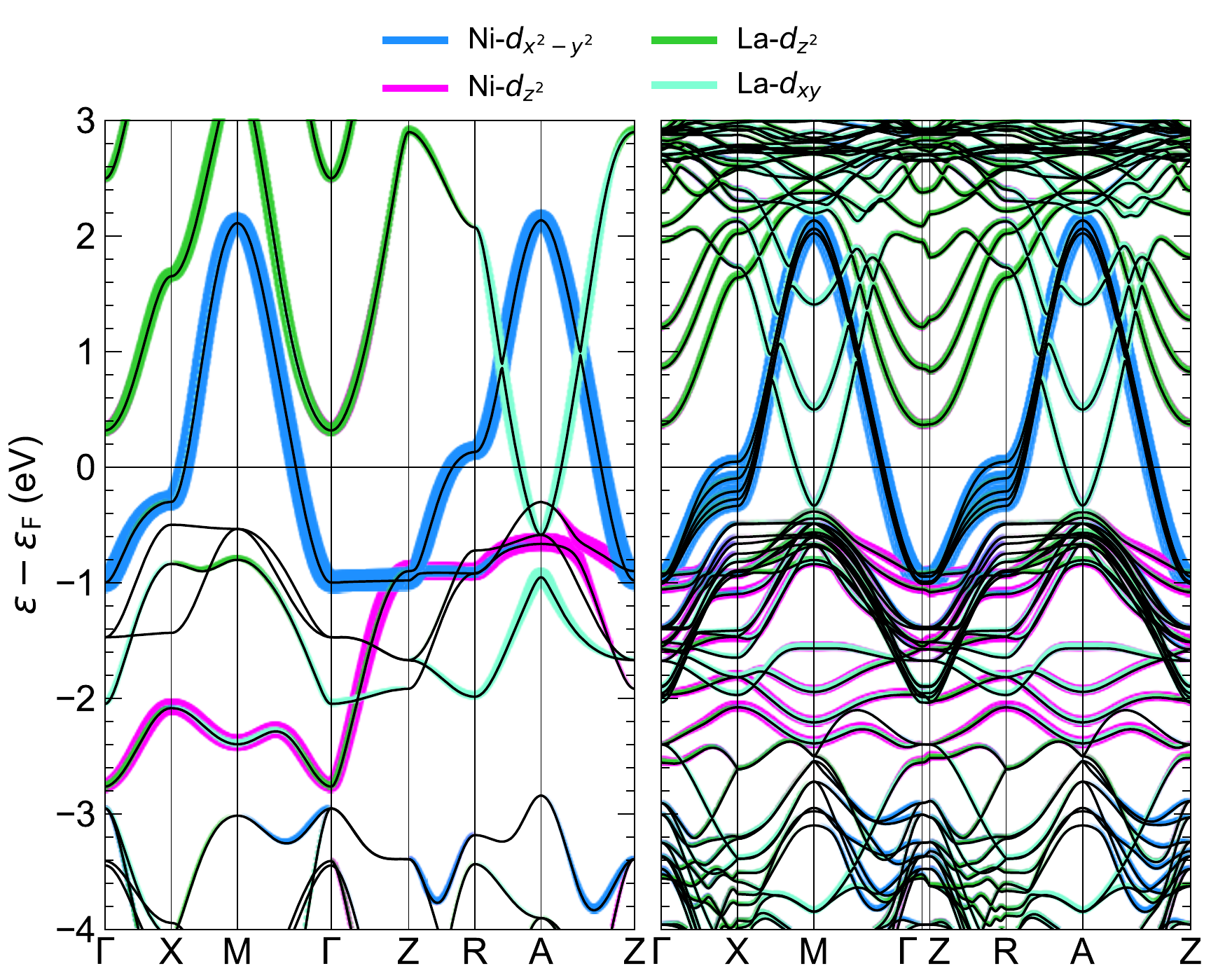}
\caption{DFT band structures for 20\% hole-doped LaNiO$_2$ ($n=\infty$) (left) and La$_6$Ni$_5$O$_{12}$ ($n=5$) (right), both at $d^{8.8}$ nominal filling. The band structures are shown along high symmetry directions in the Brillouin zone with `fatband' representation for the Ni-$d_{x^2-y^2}$, Ni-$d_{z^2}$, La-$d_{xy}$, and La-$d_{z^2}$ orbitals.}
\label{fig:dft}
\end{figure}
We subsequently map the Kohn-Sham Hamiltonian obtained within DFT onto a basis set of atomic-like orbitals within a correlated subspace ($-10$ eV to 10 eV around the Fermi energy) using the projection method provided by the TRIQS/DFTtools software package \cite{triqs_wien2k_interface, triqs_dft_tools}. Local Coulomb interactions are added to our effective Hamiltonian defined in this correlated subspace. We have chosen the Ni-$e_{g}$ \{$d_{x^{2}-y^{2}}$, $d_{z^{2}}$\} orbitals as our correlated subspace and include interactions of the Hubbard-Kanamori form,
\begin{multline}
   \mathcal{H}_{\mathrm{int}} = U \sum_{m} \hat{n}_{m\uparrow}\hat{n}_{m\downarrow} + (U-2J) \sum_{m\neq m'}\hat{n}_{m\uparrow}\hat{n}_{m'\downarrow} \\
   +(U-3J) \sum_{m<m', \sigma}\hat{n}_{m\sigma}\hat{n}_{m'\sigma} \\
   +J \sum_{m\neq m'} \hat{c}^{\dagger}_{m\uparrow}\hat{c}^{\dagger}_{m\downarrow}\hat{c}_{m'\downarrow}\hat{c}_{m'\uparrow} - J \sum_{m\neq m'}\hat{c}^{\dagger}_{m\uparrow}\hat{c}_{m\downarrow}\hat{c}^{\dagger}_{m'\downarrow}\hat{c}_{m'\uparrow},
\end{multline}
where $\hat{c}^{\dagger}_{m\sigma}$ creates an electron in the correlated atomic orbital $m$ with spin $\sigma$. We choose a local Coulomb repulsion $U$= 7 eV and Hund's coupling $J$= 0.7 eV, typical values for nickelates \cite{Karp2020_112, Karp2020_438}. The Held's double counting formula has been used \cite{Held2007}, 
\begin{equation}
    \Sigma_{\mathrm{dc}} = \frac{U + (d-1)(U-2J) + (d-1)(U-3J)}{2d-1} \big (n- \frac{1}{2} \big)
\end{equation}
where $d$ is the number of correlated orbitals and $n$ is the density of the correlated orbitals, to subtract the Hartree contribution to the self-energy that is already approximated within DFT. Single-site DMFT calculations are performed using the TRIQS software library \cite{triqs}, where the impurity problem is solved with the continuous-time hybridization expansion solver ({\sc cthyb}) \cite{cthyb} at a temperature of $T = 290$ K ($\beta$ = 40 eV$^{-1}$). To avoid high-frequency noise in the impurity self-energy and Green's function we represent both quantities in the Legendre basis and sample the Legendre coefficients directly within the TRIQS/{\sc cthyb} solver \cite{Boehnke2011legendre}.  For the 5-layer nickelate, we solve three impurity problems for the three inequivalent Ni sites in the inner, middle, and outer NiO$_{2}$ layers. Maximum entropy methods are employed for the analytical continuation from Matsubara space onto the real frequency axis \cite{triqs_maxent}. Our calculations are ``one-shot'' DFT+DMFT calculations meaning the DFT charge density is not updated. Recent studies have shown that one-shot calculations are sufficient to gain qualitative insights into the many-body electronic structure of transition-metal oxides \cite{Karp2021, Hampel2020effect}. Specifically, Ref. \onlinecite{Karp2021} showed that there are small differences between one-shot and charge self-consistent DFT+DMFT calculations for NdNiO$_{2}$. Nevertheless, we have performed careful benchmarks to ensure our one-shot calculations describe the correlated electronic structure accurately. For our benchmark studies (see Appendix \ref{sec:benchmark}), we focus on the electronic structure of the infinite-layer nickelate at $d^9$ nominal filling that has been reported in previous work \cite{Karp2020_112, Karp2020_438, Karp2021, lechermann2020late, lechermann2020multi, Wang2020, Kang2021opt}.

%%%%%%%%%%%%%%%%%%% DFT RESULTS %%%%%%%%%%%%%%%%%%%%%%%%%%%%%%%%%%%

\section{\label{sec:results}Results}

\subsection{\label{sec:dft_s}DFT electronic structure}
Figure \ref{fig:dft} displays the band structure along high symmetry directions obtained from DFT for 20\% hole-doped LaNiO$_2$ ($n=\infty$) and La$_6$Ni$_5$O$_{12}$ ($n=5$) in the paramagnetic state (both at $d^{8.8}$ nominal filling). The Ni-$d_{x^2-y^2}$ and Ni-$d_{z^2}$ orbital character of the bands is highlighted, as well as that for the La-$d_{z^2}$ and La-$d_{xy}$  orbitals. For infinite-layer LaNiO$_{2}$, the band structure shows a single Ni-$d_{x^{2}-y^{2}}$ band crossing the Fermi level (akin to cuprates), but with an extra electron pocket of La-$d_{xy}$ character appearing at A. In the parent material (at $d^{9}$ filling), there is an additional pocket of La-$d_{z^{2}}$ character appearing at $\Gamma$ \cite{Karp2020_112, botana2020, Leonov2020, Kapeghian2020, lechermann2020late, lechermann2020multi, pickett2020, labollita2021, kitatani2020, krishna2020}. The additional rare-earth band(s) give rise to a self-doping effect that has been the subject of ample scrutiny \cite{botana2020, pickett2004, krishna2020, Karp2020_112, labollita2021}. For the 5-layer nickelate La$_6$Ni$_5$O$_{12}$, there are five Ni-$d_{x^{2}-y^{2}}$ bands crossing the Fermi level (one per layer). The splitting in the Ni-$d_{x^{2}-y^{2}}$ bands at X is a consequence of the interlayer hopping, similar to the multi-layer cuprates \cite{Sakakibara2014}. Electron pockets at M and A also have a dominant La-$d_{xy}$ orbital character. All in all, the infinite-layer and 5-layer material have, when compared at the same filling, identical-character active bands crossing the Fermi energy. 
An estimate of the amount of self-doping in both materials (at $d^{8.8}$ filling) as obtained from the area of their electron-like Fermi pockets gives $\sim$ 0.023 electrons in the infinite-layer compound (from the electron pockets at A), while for the 5-layer material the pockets at M and A enclose $\sim$ 0.025 electrons. One notable difference arises when looking at the Fermi surfaces (see Appendix \ref{sec:dft_fs}): the fermiology of the 5-layer nickelate is much more two-dimensional-like and reminiscent of the multi-layer cuprates  with single sheets originating from the Ni-$d_{x^2-y^2}$ states, even though there are additional pockets at the corners of the Brillouin zone of La-$d$ character  \cite{pan2021super}. This difference in the degree of two-dimensionality with respect to the infinite-layer material arises from the structural differences described above  (more specifically it is due to the fluorite blocking slab present in the 5-layer nickelate). 
%%%%%%%%%%%%%%%%%%% FIG Im \Sigma (i\omega_{n}) %%%%%%%%%%%%%%%%%%%%%%%%%%%%%%%%%%%
\begin{figure}
	\centering
	\includegraphics[width=\columnwidth]{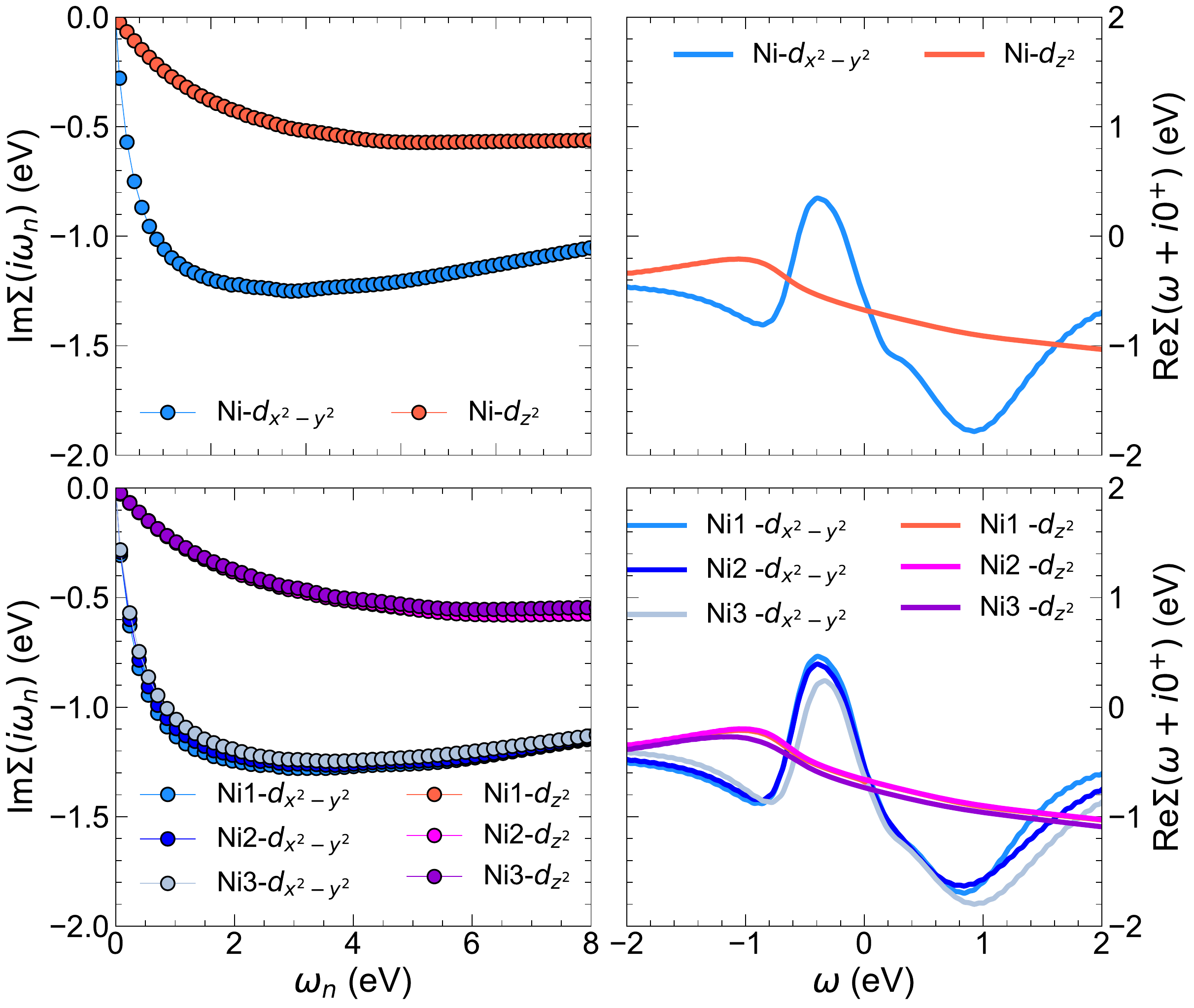}
	\caption{Electronic self-energies for 20\% hole-doped LaNiO$_2$ ($n=\infty$) (top panels) and La$_6$Ni$_5$O$_{12}$ ($n=5$) (bottom panels) both at $d^{8.8}$ filling. Left panels: $d_{x^2-y^2}$ and $d_{z^2}$ components of the imaginary part of the electron self-energy in Matsubara space for $n=\infty$ (top) and $n=5$ (bottom). Right panels: real part of the analytically continued self-energies Re$\Sigma(\omega + i0^{+})$ for the $d_{x^2-y^2}$ and $d_{z^2}$ orbitals, where the double counting correction has been subtracted from the self-energies.}
	\label{fig:sigma}
\end{figure}

In the cuprate context, the degree of hybridization between O-$p$ and Cu-$d$ orbitals is always an important quantity to consider, given that it is relevant for Zhang-Rice singlet formation \cite{zhang1988}. The degree of $p-d$ hybridization can be quantified via the charge-transfer energy, $\Delta = \varepsilon_{d} - \varepsilon_{p}$, where $\varepsilon_{d}$ and $\varepsilon_{p}$ are the  transition metal-$d$ and O-$p$ on-site energies, respectively. For cuprates, the charge-transfer energy ranges from $1-2$ eV. The estimates we obtain for the charge-transfer energies in the 20\% hole-doped $n=\infty$ and $n=5$ nickelates using the on-site energies from maximally localized Wannier functions (MLWFs) \cite{wannier90, wien2wannier} are shown in Table \ref{tab:on-site} (further details are shown in Appendix \ref{sec:wannier}). For the infinite-layer nickelate, $\varepsilon_{d}-\varepsilon_{p}$ (referring to Ni-$d_{x^2-y^2}$ and O-$p_{\sigma}$) is $\sim$ 3.9 eV, a $\sim$0.5 eV reduction with respect to the charge-transfer energy at $d^{9}$ \cite{botana2020, krishna2020}. For the $n=5$ nickelate, the charge-transfer energy is layer-dependent, averaging to nearly the same value $\sim$ 4.0 eV \cite{labollita2021}. 

\begin{table}
    \centering
    \begin{tabular*}{\columnwidth}{l@{\extracolsep{\fill}}lccc}
    \hline
    \hline
        $n$ &  NiO$_{2}$ layer & $\varepsilon_{p_{\sigma}}$ (eV) & $\varepsilon_{d_{^{x^2-y^2}}}$ (eV) & $\Delta$ (eV)\\
        \hline
        $\infty$ & --     & $-4.88$ & $-0.98$ & 3.90\\
        
        $5$      & inner  & $-4.93$ & $-0.97$ & 3.96\\
                 & middle & $-4.81$ & $-0.94$ & 3.87\\
                 & outer  & $-4.86$ & $-0.96$ & 3.78\\
        %$5$      & inner  & $-5.15$ & $-1.03$ & 4.12\\
        %         & middle & $-5.07$ & $-1.01$ & 4.06\\
        %         & outer  & $-4.86$ & $-0.98$ & 3.88\\
    \hline
    \hline
    \end{tabular*}
    \caption{On-site energies obtained from MLWFs for 20\% hole-doped LaNiO$_2$ ($n=\infty$) and La$_6$Ni$_5$O$_{12}$ ($n=5$) both at $d^{8.8}$ nominal filling (with respect to the Fermi energy). The charge-transfer energy is derived from $\Delta = \varepsilon_{d_{x^2-y^2}}-\varepsilon_{p_{\sigma}}$ for both materials. Note that $p_{\sigma}$ denotes the bonding O-$p$ orbital with the Ni-$d_{x^{2}-y^{2}}$ orbital.}
    \label{tab:on-site}
\end{table}

\begin{table}
	\centering
	\begin{tabular*}{\columnwidth}{l@{\extracolsep{\fill}}lcc}
	\hline
	\hline
	$n$     &  NiO$_{2}$ layer & $d_{x^{2}-y^{2}}$ & $d_{z^{2}}$ \\
	\hline
	$\infty$ &    --           &   3.89              & 1.25        \\ %Pro j -10 to 10 dx2y2 4.6 dx2 = 1.3 https://arxiv.org/pdf/2102.08522.pdf
	\hline
	5       & 	 inner         &   4.30              & 1.29        \\
	        &    middle        &   4.06              & 1.29       \\
	        & 	 outer         &   3.83              & 1.29        \\
	\hline 
	\hline
	\end{tabular*}
	\caption{Mass enhancements ($m^{\star}/m$) for the $d_{x^{2}-y^{2}}$ and $d_{z^{2}}$ orbitals obtained from the imaginary part of the electronic self-energy for 20\% hole-doped LaNiO$_2$ ($n=\infty$) and La$_6$Ni$_5$O$_{12}$ ($n=5$), both at $d^{8.8}$ nominal filling.}
	\label{tab:mass_enhance}
\end{table}

%%%%%%%%%%%%%%%%%%% SELF-ENERGIES%%%%%%%%%%%%%%%%%%%%%%%%%%%%%%%%%%%
\begin{figure*}
	\centering
	\includegraphics[width=1.95\columnwidth]{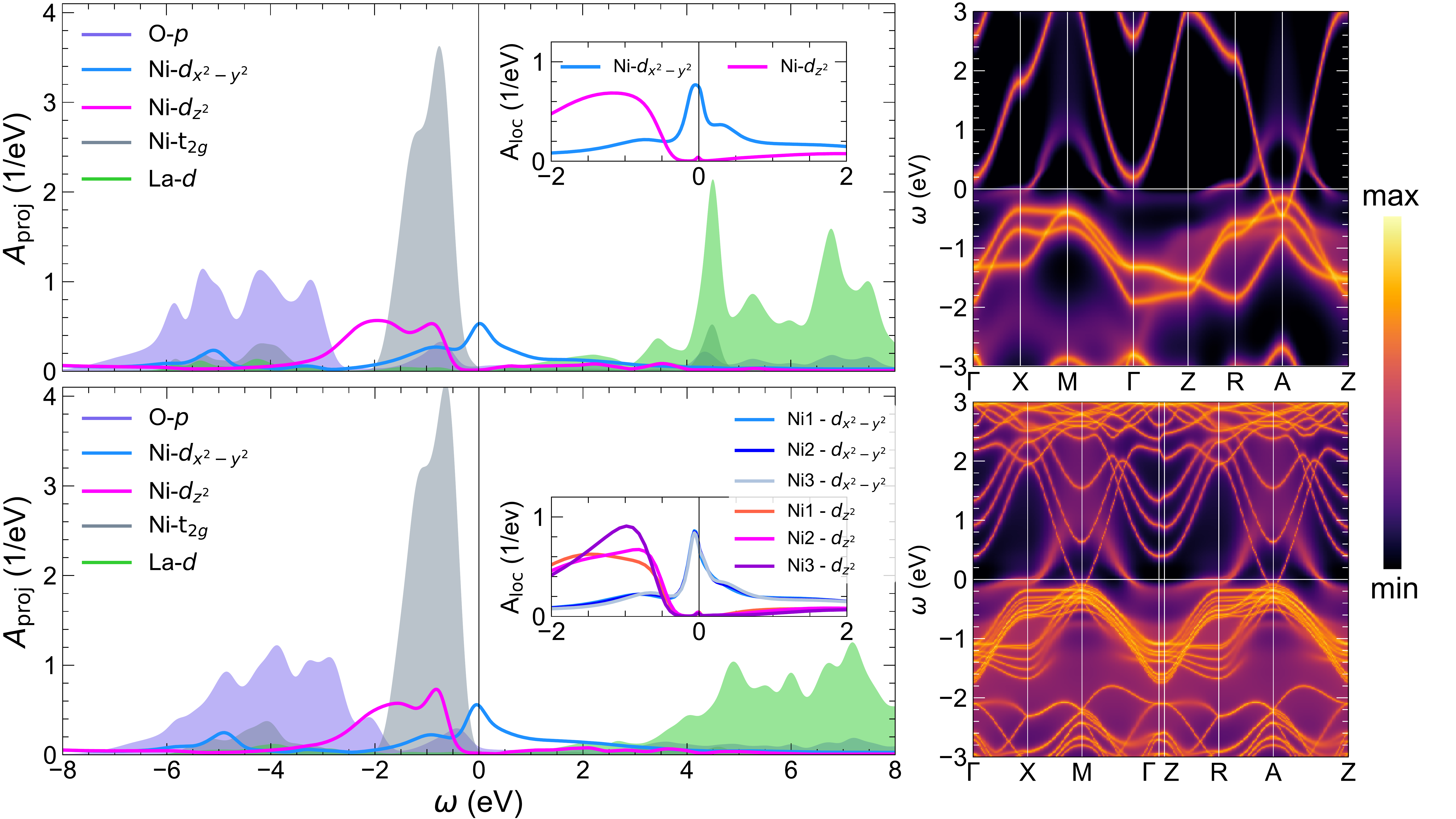}
	\caption{Spectral properties for 20\% hole-doped LaNiO$_2$ ($n=\infty$) (top panels) and La$_6$Ni$_5$O$_{12}$ ($n=5$) (bottom panels) both at $d^{8.8}$ nominal filling.  Orbital-projected spectral functions (left panels) where the inset shows the local Ni-$e_{g}$ spectral functions and $\vb{k}$-resolved spectral function $A(\vb{k}, \omega)$ along high-symmetry lines in the Brillouin zone (right panels).}
	\label{fig:spectral}
\end{figure*}

\subsection{\label{sec:corr_es}Correlated electronic structure}
\subsubsection{\label{sec:selfenergies}Self-energies}
We now turn to the many-body electronic structure. Figure \ref{fig:sigma} shows the frequency dependence of the self-energy on the imaginary and real axis for 20\% hole-doped LaNiO$_2$ ($n=\infty$) and La$_6$Ni$_5$O$_{12}$ ($n=5$), both at $d^{8.8}$ nominal filling. For both materials, the $d_{x^2-y^2}$ component has a much steeper slope in the low frequency regime compared to the $d_{z^2}$ component, indicating that the $d_{x^2-y^2}$ orbital is more strongly correlated. For the three inequivalent Ni impurity sites in the 5-layer material, the imaginary part of the self-energy in Matsubara space is similar with subtle variations in the low frequency range. Specifically, the outer Ni differs relative to the inner and middle Ni sites likely due to the different local environment of the outer Ni that has a single neighboring NiO$_{2}$ plane (this is in contrast to inner and middle planes, see Fig. \ref{fig:struct}). To quantify the strength of correlations, we calculate the mass enhancements from the inverse quasiparticle renormalization factor, $m^{\star}/m = Z^{-1}$. We obtain $Z^{-1}$ directly from the self-energies in Matsubara space to avoid any ambiguity introduced through analytic continuation \cite{Karp2020_112, Karp2020_438, mravlje2011, zingl2019}. Specifically, $Z^{-1}$ is calculated by fitting a fourth-order polynomial to the lowest Matsubara frequencies, then the renormalization factor is given by $Z^{-1} = \big (1 - \partial \mathrm{Im}\Sigma(i\omega_{n})/ \partial \omega_{n}\big |_{\omega_{n}\rightarrow 0^{+}} \big )$ \cite{mravlje2011, zingl2019}. The mass enhancements are summarized in Table \ref{tab:mass_enhance} for the two correlated orbitals ($d_{x^{2}-y^{2}}$ and $d_{z^{2}}$) for both materials. For the $n=\infty$ member, we find $m^{\star}/m \sim 3.9$ for the $d_{x^{2}-y^{2}}$ orbital and $m^{\star}/m \sim 1.25$ for the $d_{z^{2}}$ orbital. The $d_{x^{2}-y^{2}}$ mass enhancement slightly decreases at $d^{8.8}$ nominal filling compared to $d^{9}$ (see Appendix \ref{sec:benchmark}). In the 5-layer material, $m^{\star}/m$ for the $d_{x^{2}-y^{2}}$ orbitals is $\sim 3.8-4.3$ for the three inequivalent Ni sites, with some slight variations on the inner, middle, and outer layers. For the $d_{z^2}$ orbitals a much smaller mass enhancement $m^{\star}/m \sim$ 1.3 is derived. At the same carrier concentration, the mass enhancements are very similar for both materials. The mass enhancements derived above are obtained from the electronic self-energy in the basis of correlated orbitals. Another relevant quantity is the ``band basis'' mass enhancement as it provides an indication of the amount of admixture of the correlated orbitals with the uncorrelated ones. We  obtain the band basis mass enhancement by upfolding the electronic self-energy in the orbital basis using our projectors (see Appendix \ref{sec:Zbands} for more details). In the band basis, we find that the mass enhancement for the $d_{x^{2}-y^{2}}$ band near the Fermi energy decreases to around $\sim 2.8$ for the infinite-layer and to $\sim 2.3$ for the 5-layer material (see Fig. \ref{fig:band_mass} in Appendix \ref{sec:Zbands}). This large decrease in mass enhancement for both materials is an indication of the strong hybridization between the Ni-$d$ and O-$p$ orbitals.  Overall, our results confirm that the correlations in this family of layered nickelates are dominated by the $d_{x^2-y^2}$ orbitals. 

From the real part of the analytically continued self-energy, we find that that the $d_{x^{2}-y^{2}}$ self-energy has substantial particle-hole symmetric structures around $\omega=0$, a consequence of the Mott-Hubbard and charge-transfer correlations \cite{Karp2020_438}. The $d_{z^{2}}$ self-energy is much smoother around $\omega=0$ indicative of weaker correlations. Importantly, the structure in the $d_{x^{2}-y^{2}}$ self-energy is essentially identical between the two materials. The size of these structures has a dependence on carrier concentration and becomes more pronounced at $d^{9}$ \cite{Karp2020_438} (see Appendix \ref{sec:benchmark}). This indicates a weakening of correlations upon hole-doping expected of Mott-Hubbard/charge-transfer materials. In the 5-layer compound, the outer Ni seems to display a slightly different self-energy with respect to the other two Ni atoms, likely due to the difference in environment already highlighted above.

%%%%%%%%%%%%%%%%%%% SPECTRAL FUNCTIONS %%%%%%%%%%%%%%%%%%%%%%%%%%%%%%%%%%%
\subsubsection{\label{sec:spectral}Spectral functions}
Figure \ref{fig:spectral} summarizes the spectral properties for 20\% hole-doped LaNiO$_2$ ($n=\infty$) and La$_6$Ni$_5$O$_{12}$ ($n=5$), both at $d^{8.8}$ nominal filling. The orbital-resolved spectral function defined as $A(\omega) = \frac{i}{2\pi}(G(\omega) -G^{\dagger}(\omega))$ is the interacting analog to the DFT density of states (DOS). We find that the spectral functions are qualitatively and quantitatively similar between the two nickelates and agree well with the DOS calculated within DFT \cite{labollita2021}. In the addition spectrum ($\omega > 0$), the La-$d$ states seem to be located at the same energy for both systems.  Note that at $d^9$ for the parent infinite-layer nickelate, the La-$d$ states are closer to the chemical potential, that is, shifted down to lower energies (see Appendix \ref{sec:benchmark}). In the removal spectrum ($\omega < 0$), we see that the centroid of the O-$p$ states is located at the same energy in both materials.  The Ni-$t_{2g}$ and Ni-$e_{g}$ states are essentially fixed between the two nickelates as well, with the Ni-$d_{x^{2}-y^{2}}$ states being the dominant ones around the Fermi energy. The similar charge transfer
energies discussed above can be visualized qualitatively here as the energy separation between the peaks in the Ni-$d$ and O-$p$ projected spectral functions, which do not seem to differ between the two materials. The local spectral functions (insets in Fig. \ref{fig:spectral}) are obtained through analytic continuation of the impurity Green's function. We find that the features in the local $e_{g}$ spectral functions are essentially the same for both materials with the characteristic three-peak structure in the $d_{x^{2}-y^{2}}$ component, which corresponds to a central quasiparticle peak near the chemical potential with lower and upper Hubbard bands. For the 5-layer nickelate, the inner and middle-Ni impurity sites exhibit nearly identical local spectral functions, while the outer impurity shows some variation. Specifically, there is a subtle difference in the $d_{z^{2}}$ component of the spectral function with a much stronger peak in the removal spectra. We attribute this difference once again to the different local environment of the outer layer nickel.

% occupations and probablities for dx2y2 and z2
%https://arxiv.org/pdf/2102.08522.pdf ndx2y2= 1.14, ndz2 = 1.65, N = 2 (LS) 0.11 , N = 2 (HS) 0.15, N = 3 0.64, N = 4 0.09
% occupations for non-correlated orbitals
%https://arxiv.org/pdf/2001.06441.pdf O-p 3.60 Nd 0.53
\begin{table*}
\begin{tabular*}{1.8\columnwidth}{l@{\extracolsep{\fill}}lcccccccc}
\hline
\hline
$n$      &  NiO$_{2}$ layer & Ni-$d_{x^{2}-y^{2}}$ & Ni-$d_{z^{2}}$ & O-$p$ & La-$d$ & $d^{8}$ & $d^{9}$ ($d_{x^{2}-y^{2}}$) & $d^{9}$ ($d_{z^{2}}$) & $d^{10}$\\
\hline
$\infty$ &       --         &   1.11               & 1.69            &  3.45 & 0.32 & 0.26 	    & 0.55                        &  0.09	               & 0.09\\
\hline
5         &  inner          &   1.12               & 1.69            &   --  & --   & 0.25 	    & 0.56                        &  0.09	               & 0.09\\
          &  middle         &   1.11               & 1.69            &   --  & --   & 0.26 	    & 0.55                        &  0.09 	               & 0.09\\
          &  outer          &   1.10               & 1.72            &   --  & --   & 0.25      & 0.57                        &  0.08  	               & 0.09\\
          &    --           &     --               &    --           &  3.55 & 0.30 & --       & -- & -- & -- \\
\hline
\hline
\end{tabular*}
\caption{Left: Orbital-resolved occupancies obtained from the impurity Green's function $G(i\omega_{n})$ for 20\% hole-doped LaNiO$_2$ ($n=\infty$) and La$_6$Ni$_5$O$_{12}$ ($n=5$), both at $d^{8.8}$ nominal filling. Right: Occurrence probabilities for different Ni $d$ valence states obtained from the impurity density matrices for the two materials.}
\label{tab:occ}
\end{table*}

\begin{figure}
    \centering
    \includegraphics[width=\columnwidth]{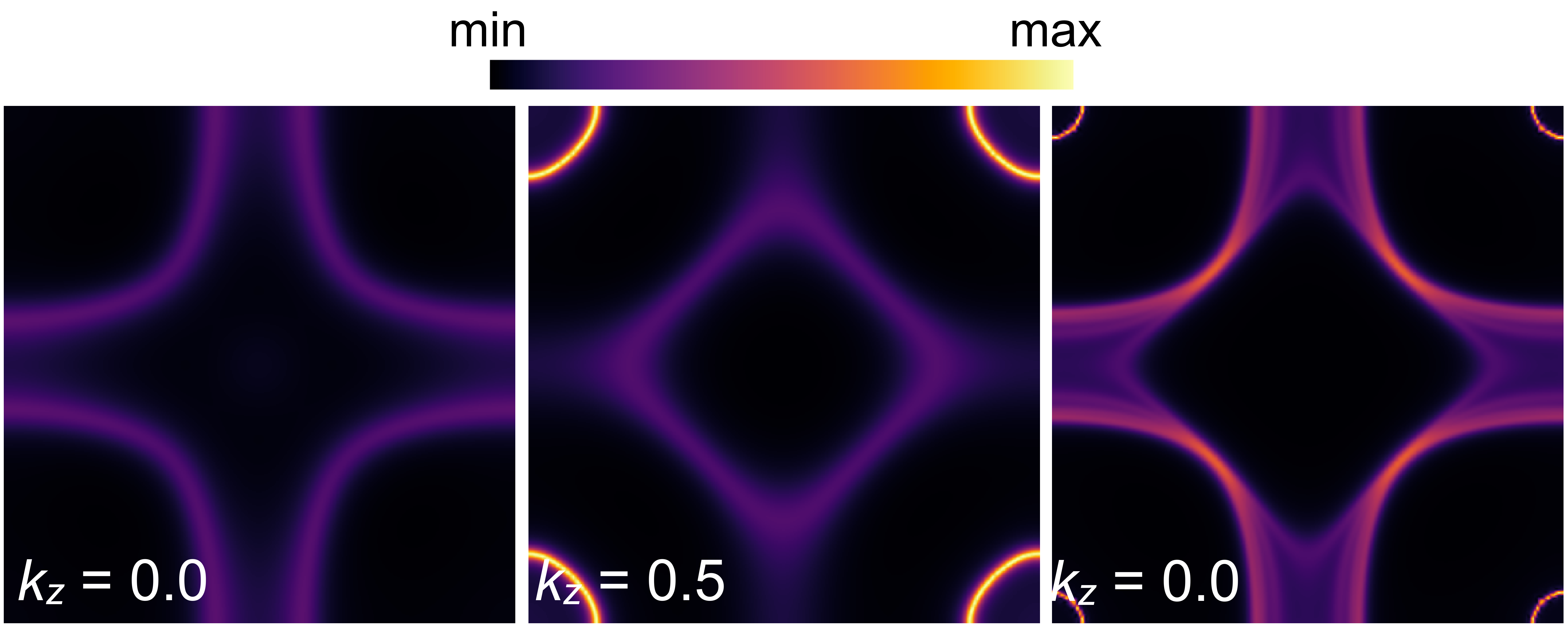}
    \caption{Interacting Fermi surfaces $A(\vb{k}, \omega=0)$ for 20\% hole-doped LaNiO$_2$ ($n=\infty$) (left, middle panels) and La$_6$Ni$_5$O$_{12}$ ($n=5$) (right panel) both at $d^{8.8}$ nominal filling. For the $n=\infty$, the Fermi surface is shown in the $k_{z}=0$ and $k_{z}=0.5$ planes showcasing the three-dimensionality of the Fermi surface compared to the $n=5$ material.}
    \label{fig:A_fs}
\end{figure}

The momentum-resolved spectral functions, $A(\vb{k}, \omega) = - \frac{1}{\pi}\mathrm{Tr}[\mathrm{Im}G(\vb{k}, \omega)]$ along high-symmetry directions in the Brillouin zone for 20\% hole-doped LaNiO$_2$ ($n=\infty$) and La$_6$Ni$_5$O$_{12}$ ($n=5$) at $d^{8.8}$ nominal filling are also shown in Fig. \ref{fig:spectral}. The many-body electronic structure is well represented as a set of bands renormalized from the DFT values by correlations and it exhibits many of the same qualitative features for both compounds: Ni-$d_{x^{2}-y^{2}}$ band(s) with additional La-$d$ pockets crossing the Fermi level, the latter giving rise to the aforementioned self-doping effect which is absent in the cuprates. The key difference between these materials electronically is the $c$-axis dispersion: in the infinite-layer material, there is a highly-dispersive Ni-$d_{z^{2}}$ band from $\Gamma-\mathrm{Z}$ indicating a strong bonding between the NiO$_{2}$ layers. However, in the 5-layer material the $c$-axis dispersion is suppressed due to the presence of the fluorite block described previously that makes the coupling between the 5-NiO$_{2}$ blocks weak. Finally, the interacting Fermi surfaces shown in Fig. \ref{fig:A_fs} reflect how the dimensionality of the fermiology is reduced from three-dimensional in the infinite-layer case to two-dimensional in the 5-layer case, also as a consequence of the fluorite blocks present in $n \neq\infty$ layered-nickelates.

%%%%%%%%%%%%%%%%%%% ORBITAL OCCUPANICES AND OCCURENCE PROBABLITIES %%%%%%%%%%%%%%%%%%%%%%%%%%%%%%%%%%%
\subsubsection{\label{sec:orb_occu}Orbital Occupancies and Occurrence Probabilities} 
To gain further insights into the low-energy physics, we consider the relevant low-energy states for 20\% hole-doped LaNiO$_2$ ($n=\infty$) and La$_6$Ni$_5$O$_{12}$ ($n=5$)  more quantitatively. In Table \ref{tab:occ}, we have summarized the orbital-resolved occupation for the correlated orbitals, as well as mean occupations obtained from the integration of the corresponding diagonal parts of $A(\omega)$ in the projector basis over negative energies. At the same carrier concentration, the occupations of the correlated orbitals are identical for both materials with $\sim 1.7$ for the $d_{z^{2}}$ orbital and $1.1$ for the $d_{x^{2}-y^{2}}$ orbital. For the 5-layer material, across the three inequivalent Ni sites, there are only slight differences in occupation (see Table \ref{tab:occ}). We also find similar occupations of the O-$p$ orbitals in the NiO$_{2}$ planes for both materials. Importantly, the number of electrons in the La-$d$ orbitals is essentially the same at $d^{8.8}$ filling. At $d^9$ nominal filling, the occupation of the La-$d$ states increases from $\sim 0.3$ to $\sim 0.4$.
This indicates a decrease in the hybridization between the La-$d$ and Ni-$d$ states with hole doping and minimizes the relevance of the rare-earth states in the low-energy physics of the 5-layer and hole-doped infinite-layer material. This conclusion matches experimental Hall data for the Nd-based quintuple-layer nickelate wherein the Hall coefficient is positive at all temperatures, indicating that the Ni-$d$ states are the dominant low-energy states \cite{pan2021super}. In the infinite-layer nickelate, the Hall coefficient is also positive at low temperature \cite{Li2019}. In addition, previous work has shown that the degree of hybridization between R-$d$ and Ni-$d$ states in the infinite-layer materials is small, with the R-$d$ states simply acting as a charge reservoir \cite{Karp2020_112}.

To conclude, we analyze the multiplet occurrence probabilities obtained from the impurity density matrix for 20\% hole-doped LaNiO$_2$ ($n=\infty$) and La$_6$Ni$_5$O$_{12}$ ($n=5$), both at $d^{8.8}$ nominal filling. We have summarized our results in Table \ref{tab:occ}. Here, we again find essentially identical multiplet structures. For both the hole-doped infinite-layer and 5-layer material, the most probable configurations ($\sim 55$\%) correspond to $d^{9}$ Ni. The next most probable configurations are $d^{8}$ at $\sim24$\%, then $d^{10}$ at $\sim10$\%. The majority of the $d^{8}$ weight corresponds to eigenstates with high spin ($S = 1$), in agreement with previous DMFT work \cite{Karp2020_112, Karp2020_438, Karp2021, Wang2020, Kang2021opt}. We note that recent experiments in hole-doped infinite-layer nickelates \cite{Rossi2021} show that the doped holes reside mainly in the Ni-$d_{x^{2}-y^{2}}$ and are in a low-spin state, which is also supported by our DFT calculations \cite{krishna2020}. For a one band Mott-Hubbard system, one expects equal weights for $d^{10}$ and $d^{8}$ for the nominal $d^{9}$ filling. If there were more $d^{10}$ than $d^{8}$, then there would be larger charge-transfer from the oxygen orbitals (small charge-transfer energy) analogous to the cuprates. Here, more $d^{8}$ than $d^{10}$ indicates a reverse charge-transfer from the Ni-$3d$ to La-$5d$ states in both materials, such that the La-$d$ states play the role of a charge reservoir in the low-energy physics of these nickelates, as mentioned above.

%%%%%%%%%%%%%%%%%%% SUMMARY %%%%%%%%%%%%%%%%%%%%%%%%%%%%%%%%%%%
\section{\label{sec:summary}Summary}
We have employed a DFT+DMFT computational framework to compare the electronic structure of the two superconducting members of the layered rare-earth nickelate family (R$_{n+1}$Ni$_{n}$O$_{2n+1}$) with $n=\infty$ and $n=5$ at the same ($d^{8.8}$) filling. Overall, these two materials exhibit nearly identical features in their DFT and many-body electronic structure with the $d_{x^{2}-y^{2}}$ being the dominant correlated orbital while the rare-earth states near the chemical potential for both materials act as a charge reservoir. We find quantitative agreement in most aspects of the electronic structure of the two materials when comparing them at the same filling, an observation likely consistent with the fact that they exhibit nearly the same T$_c$. The most relevant difference between the two compounds is a consequence of the presence of fluorite slabs in the 5-layer nickelate that block the $c$-axis dispersion and makes the electronic structure of this material more 2D-like than that of its infinite-layer counterpart (even at the same doping). As such, the $n=5$ nickelate is more cuprate-like without the need for chemical doping. Overall, our results highlight the importance of studying layered nickelate materials at the same nominal filling to make meaningful electronic structure comparisons. Based on our findings, we note that the $n=6$ nickelate R$_7$Ni$_6$O$_{14}$ (with an average $d^{8.83}$ filling) could be an excellent candidate material to pursue to realize the next superconducting member of the layered nickelate series. 

{\it Note added.} After completion of this work, a preprint appeared \cite{worm2021correlations} reporting the many-body electronic structure of Nd$_{6}$Ni$_{5}$O$_{12}$, showing similar trends to those we present.

\section*{Acknowledgements} 
We thank J. Karp, A. J. Millis, and M. R. Norman for helpful discussions.  We acknowledge the support from NSF-DMR 2045826 and from the ASU Research Computing Center for HPC resources.

\appendix

\section{\label{sec:benchmark} DFT+DMFT calculations of LaNiO$_{2}$ at $d^{9}$ filling}
We investigate the effects of charge self-consistency on our DMFT results using the parent infinite-layer material as a benchmark, given that this material has been intensively studied in the literature \cite{Karp2020_112, Karp2020_438, Karp2021, lechermann2020late, lechermann2020late, lechermann2020multi, Kang2021opt, Wang2020}. Using the same methodology described in Sec. \ref{sec:theory}, we perform both one-shot (OS) and charge self-consistent (CSC) DFT+DMFT calculations for LaNiO$_2$ ($n=\infty$) at nominal $d^{9}$ filling. 

\begin{figure}
    \centering
    \includegraphics[width=\columnwidth]{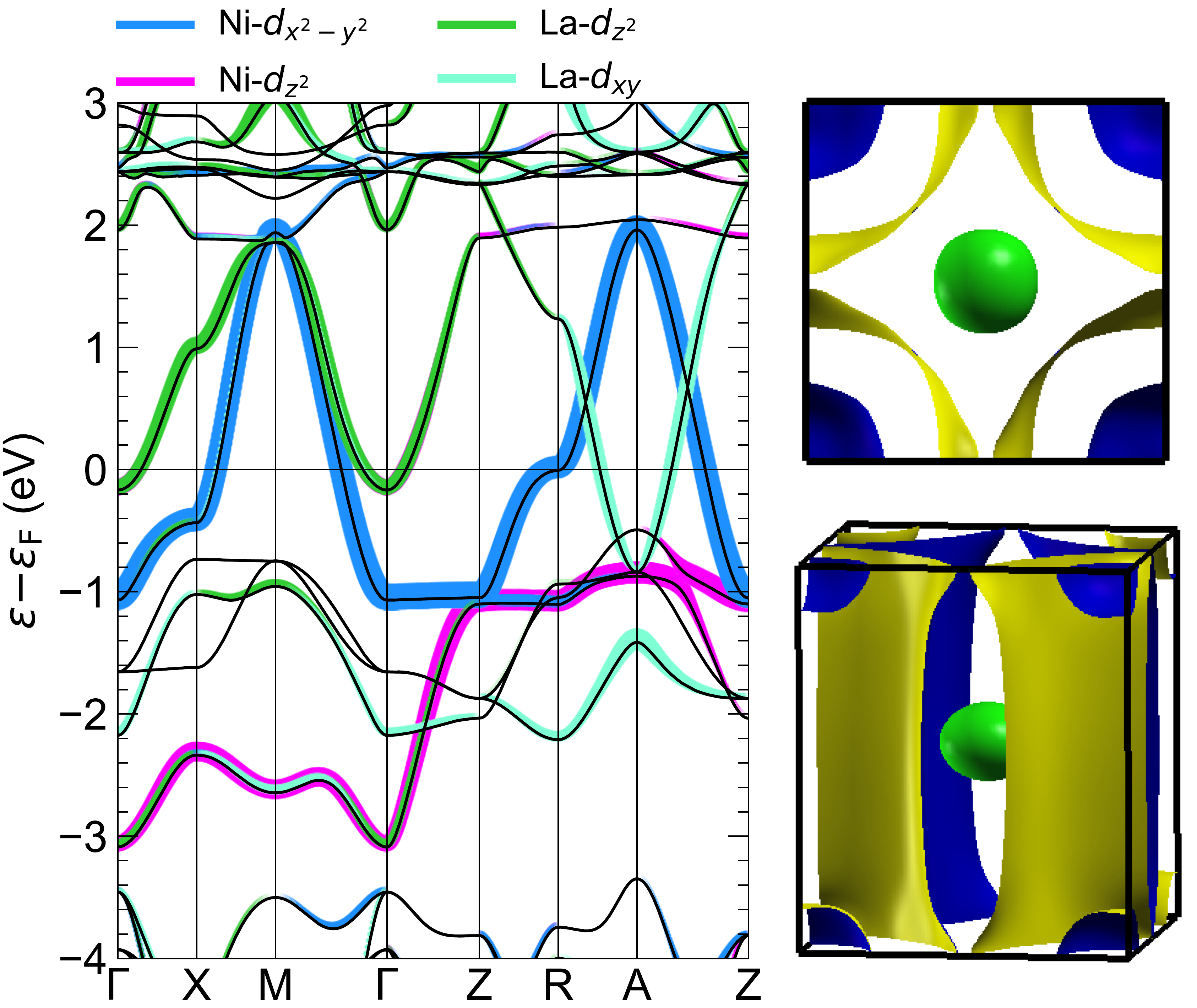}
    \caption{DFT electronic structure of parent LaNiO$_2$ (at $d^{9}$ filling). Left panel: Band structure along high-symmetry directions in the Brillouin zone with `fatband' representation for the Ni-$d_{x^2-y^2}$, Ni-$d_{z^2}$, La-$d_{xy}$, and La-$d_{z^2}$ orbitals. Right panels: Corresponding Fermi surface shown from two different perspectives: in the $k_{z}=0$ plane (top) and 3D view (bottom).}
    \label{fig:112_d9_dft}
\end{figure}

\begin{figure}
\centering
\includegraphics[width=\columnwidth]{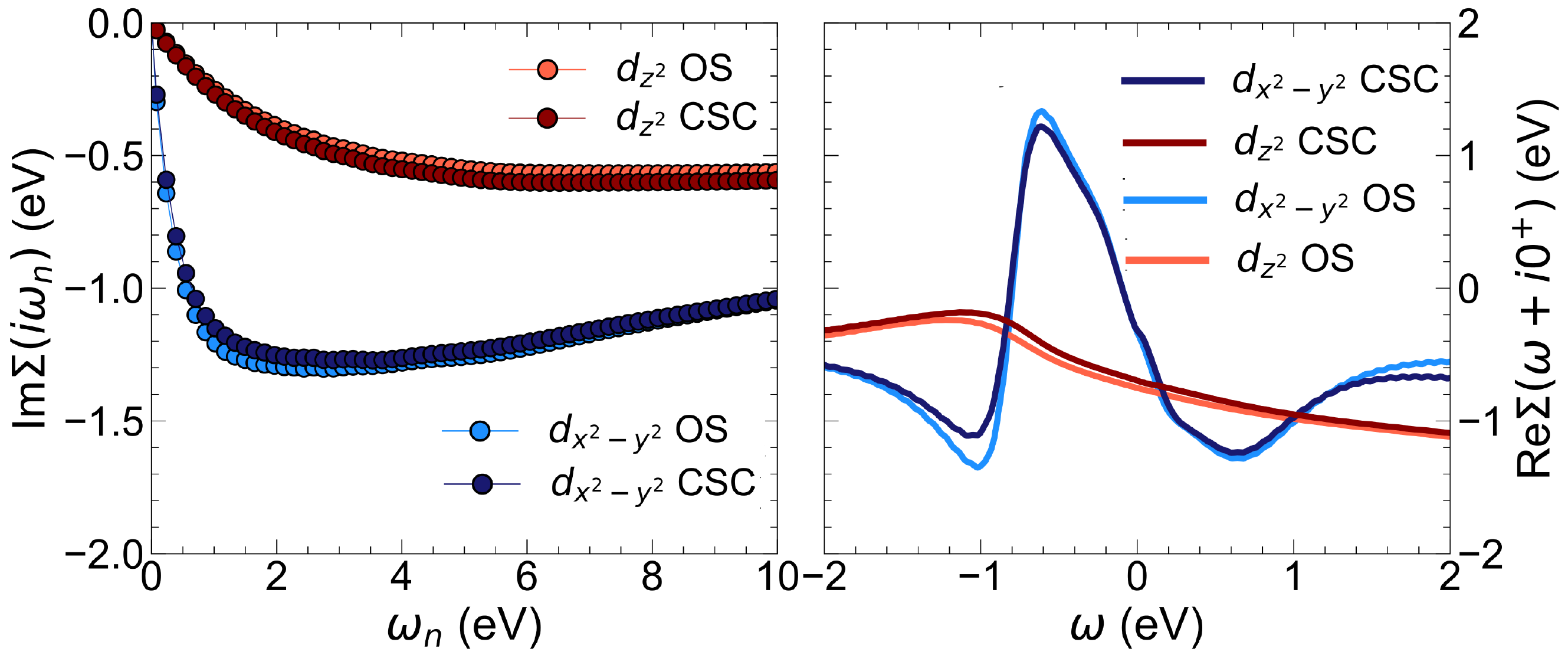}
\caption{Comparison of one-shot (OS) and charge self-consistent (CSC) DMFT calculations for parent LaNiO$_2$ ($n=\infty$) at $d^{9}$ filling. Left panel:  $d_{x^2-y^2}$ and $d_{z^2}$ components of the imaginary part of the electronic self-energy in Matsubara space. Right panel: real part of the analytically continued self-energy.}
\label{fig:112cmp}
\end{figure}
Figure \ref{fig:112_d9_dft} displays the DFT band structure along high symmetry directions for parent LaNiO$_2$ in the paramagnetic state. We highlight the orbital content of the bands around the Fermi energy, which correspond to the Ni-$d_{x^2-y^2}$, Ni-$d_{z^2}$, La-$d_{z^2}$, and La-$d_{xy}$ orbitals. The band structure we obtain for the parent infinite-layer material has been intensively described in previous literature: a single Ni-$d_{x^{2}-y^{2}}$ band crosses the Fermi level (akin to cuprates), but with two extra electron pockets of La-$d_{z^2}$ and La-$d_{xy}$ character appearing at $\Gamma$ and A, respectively \cite{Karp2020_112, botana2020, Leonov2020, Kapeghian2020, lechermann2020late, lechermann2020multi, pickett2020, labollita2021}. Additionally, we show the corresponding Fermi surface of this material containing a large hole-like sheet arising from the Ni-$d_{x^{2}-y^{2}}$ band with two electron pockets: one at $\Gamma$ with La-$d_{z^{2}}$ character, and one at A with La-$d_{xy}$ character. The Fermi surface is 3D-like due to the strong $c$-axis dispersion (see the Ni-$d_{z^{2}}$ band between $\Gamma$-Z). We note the additional electron pocket (green sphere) at $\Gamma$ is absent in the infinite-layer material at $d^{8.8}$ filling (as described in the main text) decreasing the amount of self-doping and bringing the electronic structure of the hole-doped infinite-layer nickelate closer to that of the 5-layer compound.

\begin{figure*}[ht]
\includegraphics[width=1.95\columnwidth]{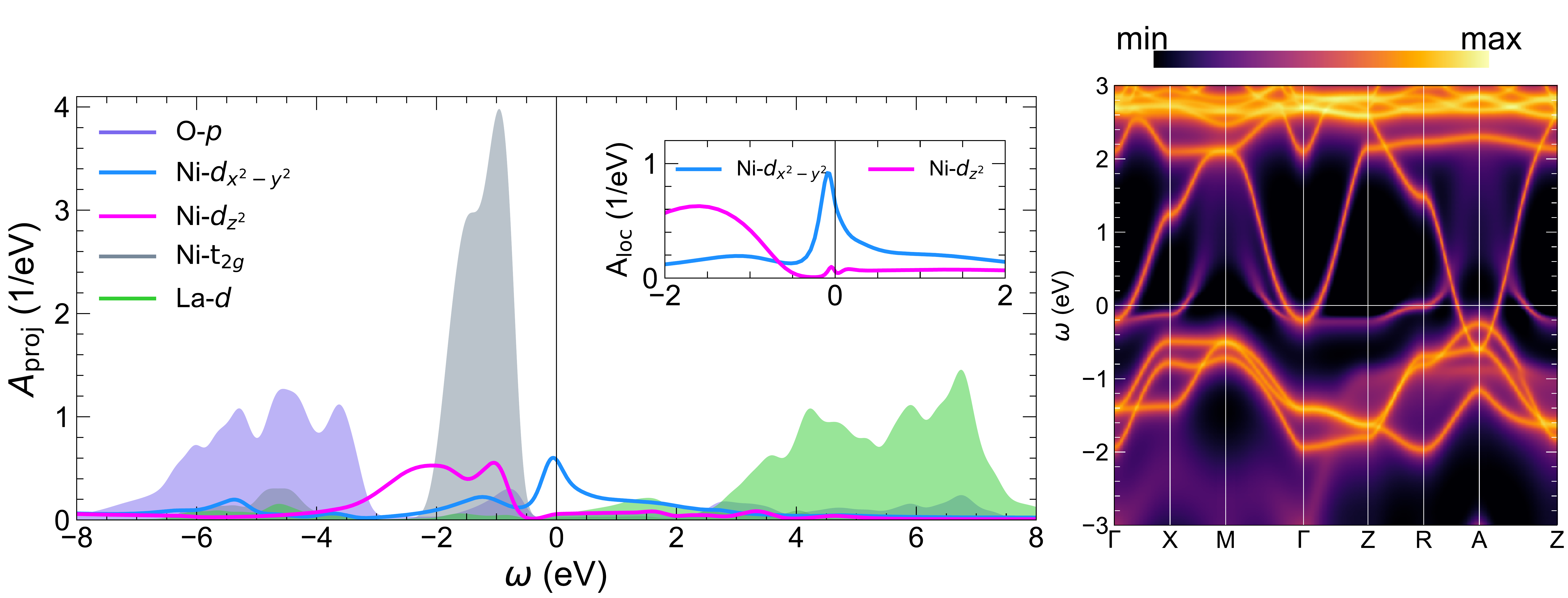}
\caption{Spectral properties of the parent 112 ($d^{9}$). Orbital-projected spectral function (left) where the inset shows the local Ni-$e_{g}$ spectral functions and $\vb{k}$-resolved spectral function $A(\vb{k}, \omega)$ along high-symmetry lines in the Brillouin zone (right).}
\label{fig:112_d9_spec}
\end{figure*}

We now compare the electronic self-energies obtained from the OS and CSC DFT+DMFT calculations. Figure \ref{fig:112cmp} shows the imaginary part of the electronic self-energy in Matsubara space. We see that both components of the self-energy are similar between the two methods, as previously shown in Ref. \cite{Karp2021}. After analytic continuation, we find that the subtle differences in the self-energies obtained from our calculations do not significantly change the structure of the self-energies on the real axis (see Fig. \ref{fig:112cmp}). Note that the particle-hole symmetric structures around $\omega=0$ are larger at $d^{9}$ compared to $d^{8.8}$ filling, indicating a weakening of correlations upon hole-doping \cite{Karp2020_438}.

For a more quantitative comparison, we calculate the mass enhancements, orbital occupancies, and occurrence probabilities, which are summarized in Table \ref{tab:112cmp}. While there are some small quantitative differences, the CSC results are very similar to the OS results. We can then conclude that charge self consistency is not crucial for our description of the many-body electronic structure of these layered nickelates. Therefore, we proceed using our OS DFT+DMFT framework throughout.
\begin{table*}
\begin{tabular*}{1.8\columnwidth}{c@{\extracolsep{\fill}}ccccccccc}
\hline
\hline
Method & $m^{\star}/m$ ($d_{x^{2}-y^{2}}$) & $m^{\star}/m$ ($d_{z^{2}}$) & $n_{d_{x^{2}-y^{2}}}$ & $n_{d_{z^{2}}}$ & $d^{8}$ & $d^{9}$ ($d_{x^{2}-y^{2}}$) & $d^{9}$ ($d_{z^{2}}$) & $d^{10}$\\
\hline
OS  & 4.29 & 1.31 & 1.17 & 1.67 & 0.24 & 0.55 & 0.11 & 0.10\\
CSC & 4.00 & 1.36 & 1.18 & 1.64 & 0.25 & 0.52 & 0.12 & 0.10\\
\hline
\hline
\end{tabular*}
\caption{Comparison of the effect of charge self consistency on the mass enhancements, orbital occupancies, and occurrence probabilities for the infinite-layer material LaNiO$_2$ at $d^9$ filling.}
\label{tab:112cmp}
\end{table*}

We summarize the corresponding spectral properties of the parent infinite-layer material in Fig. \ref{fig:112_d9_spec}. From the orbital-projected spectral function, we see that the Ni-$d_{x^{2}-y^{2}}$ states remain the dominant states around the chemical potential ($\omega=0$). Comparing to the hole-doped infinite-layer material (at $d^{8.8}$ filling), in the removal spectrum, the O-$p$ states have shifted away from the chemical potential, which increases the charge-transfer energy, while in the addition spectrum, the La-$d$ states shift closer to the chemical potential.

The $\vb{k}$-resolved spectral function along high-symmetry lines in the Brillouin zone for the parent infinite-layer material exhibits bands renormalized by correlations with respect to the DFT ones. The many-body electronic structure still exhibits many of the features of the DFT bands: a single Ni-$d_{x^{2}-y^{2}}$ band crossing the Fermi level with two additional La-$d$ pockets also crossing giving rise to a self-doping effect which is absent in the cuprates. The main difference between the parent material at $d^{9}$ and the hole-doped material at $d^{8.8}$ filling around the Fermi level is the electron pocket of La-$d_{z^{2}}$ character, which is present at $d^{9}$ and absent at $d^{8.8}$. Removing this pocket seems to bring the electronic structure of the hole-doped compound much closer to that of the 5-layer material, which also has a nominal $d^{8.8}$ filling.

\section{\label{sec:dft_fs}Fermi surfaces within DFT for infinite-layer and quintuple-layer nickelates at $d^{8.8}$ filling}
\begin{figure}
    \centering
    \includegraphics[width=\columnwidth]{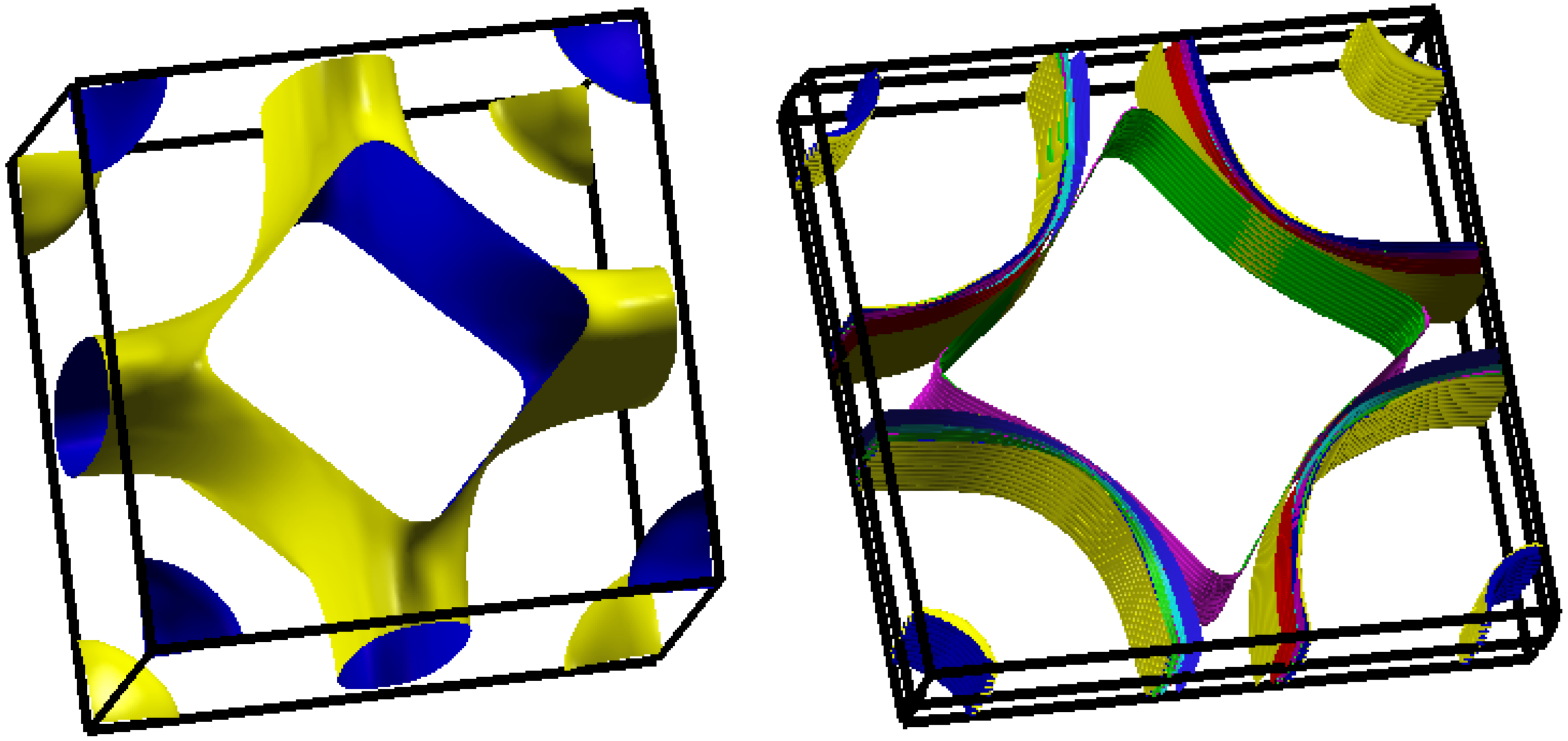}
    \caption{DFT Fermi surfaces for 20\% hole-doped LaNiO$_2$ ($n=\infty$) (left) and La$_6$Ni$_5$O$_{12}$ ($n=5$) (right) (both at $d^{8.8}$ filling). Ref. \onlinecite{labollita2021} shows Fermi surfaces for other layered nickelates of the family.}
    \label{fig:dft_fs}
\end{figure}

Figure \ref{fig:dft_fs} shows the Fermi surfaces obtained within DFT for the infinite-layer and 5-layer nickelates (both at $d^{8.8}$ nominal filling). For the infinite-layer material, the Fermi surface is 3D-like showing spherical electron pockets at the A point with dominant La-$d_{xy}$ orbital character while the hole-like sheet has Ni-$d_{x^{2}-y^{2}}$ character. For the 5-layer compound, 5 sheets corresponding to the 5 Ni-$d_{x^{2}-y^{2}}$ bands can be observed (four hole-like and one electron-like). Additionally, the electron-like pockets that can be observed at the zone corner, with La-$d_{xy}$ character, are cylindrical in the 5-layer compound, rather than spherical as in the infinite-layer compound. This latter difference in the fermiology arises from the different symmetries of the two crystal structures (see Sec. \ref{sec:crystal}). Overall, the Fermi surface of the 5-layer system is much more two-dimensional-like due to the presence of the fluorite slab which cuts off the $c$-axis dispersion. 

\section{\label{sec:wannier}Wannierizations}
\begin{figure}
\centering
\includegraphics[width=\columnwidth]{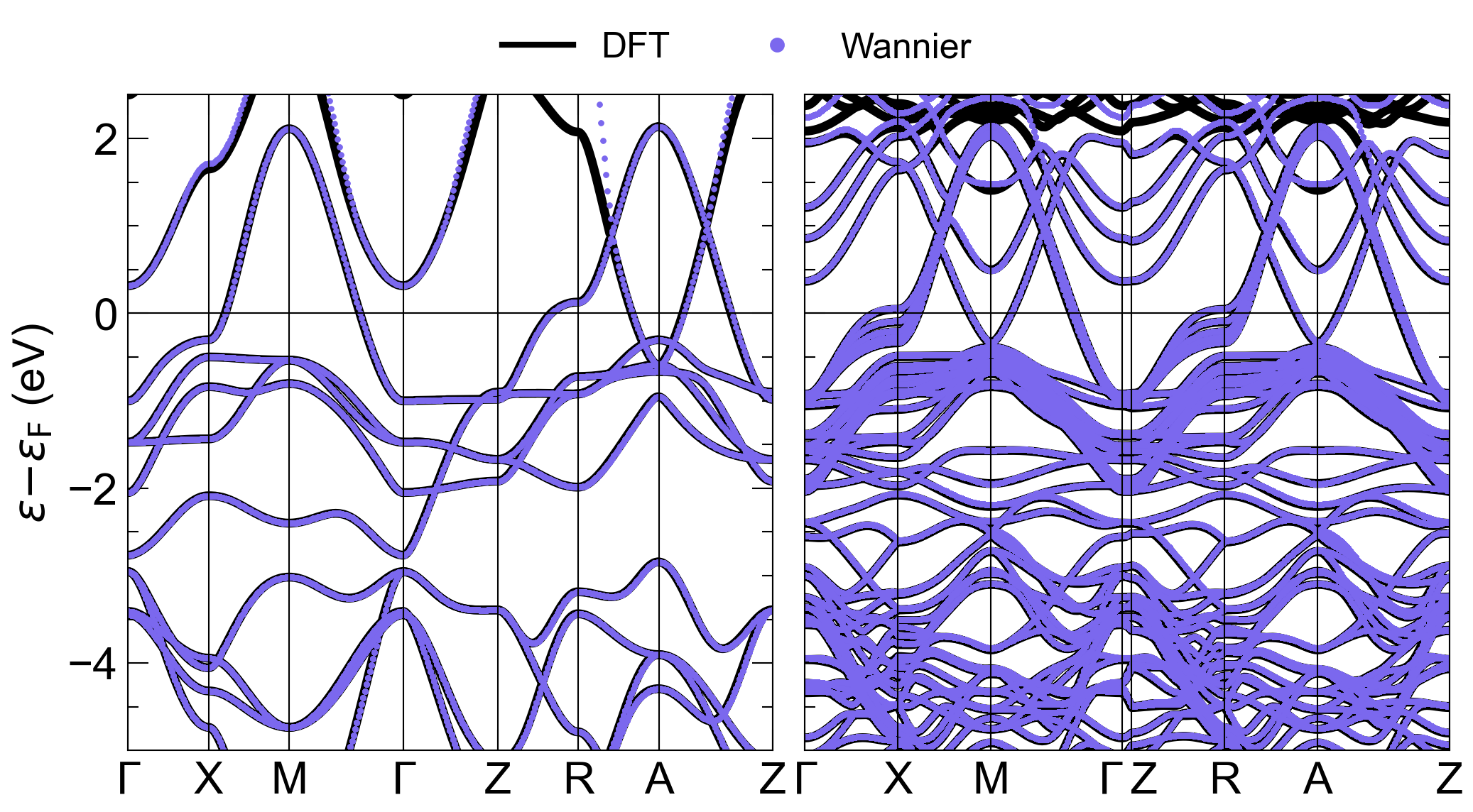}
    \caption{Wannier bands for 20\% hole-doped LaNiO$_2$ ($n=\infty$) (left) and La$_6$Ni$_5$O$_{12}$ ($n=5$) (right) (both at $d^{8.8}$ nominal filling) compared to the DFT bands.}
    \label{fig:wannier}
\end{figure}

To derive the on-site energies for an estimate of the charge-transfer energy, we obtain maximally-localized Wannier functions (MLWFs) for both LaNiO$_2$ and La$_6$Ni$_5$O$_{12}$ (at $d^{8.8}$ filling) using {\sc wannier}90  \cite{wannier90} and {\sc wien}2{\sc wannier} \cite{wien2wannier}. For both materials, we used the Ni-$d$, O-$p$, La-$d_{xy}$, and La-$d_{z^{2}}$ orbitals for our initial projections to obtain well-localized (albeit not unique) Wannier functions that correctly reproduce the band structure (see Fig. \ref{fig:wannier}).

\section{\label{sec:Zbands}Band basis mass enhancements}

\begin{figure}[h]
    \centering
    \includegraphics[width=\columnwidth]{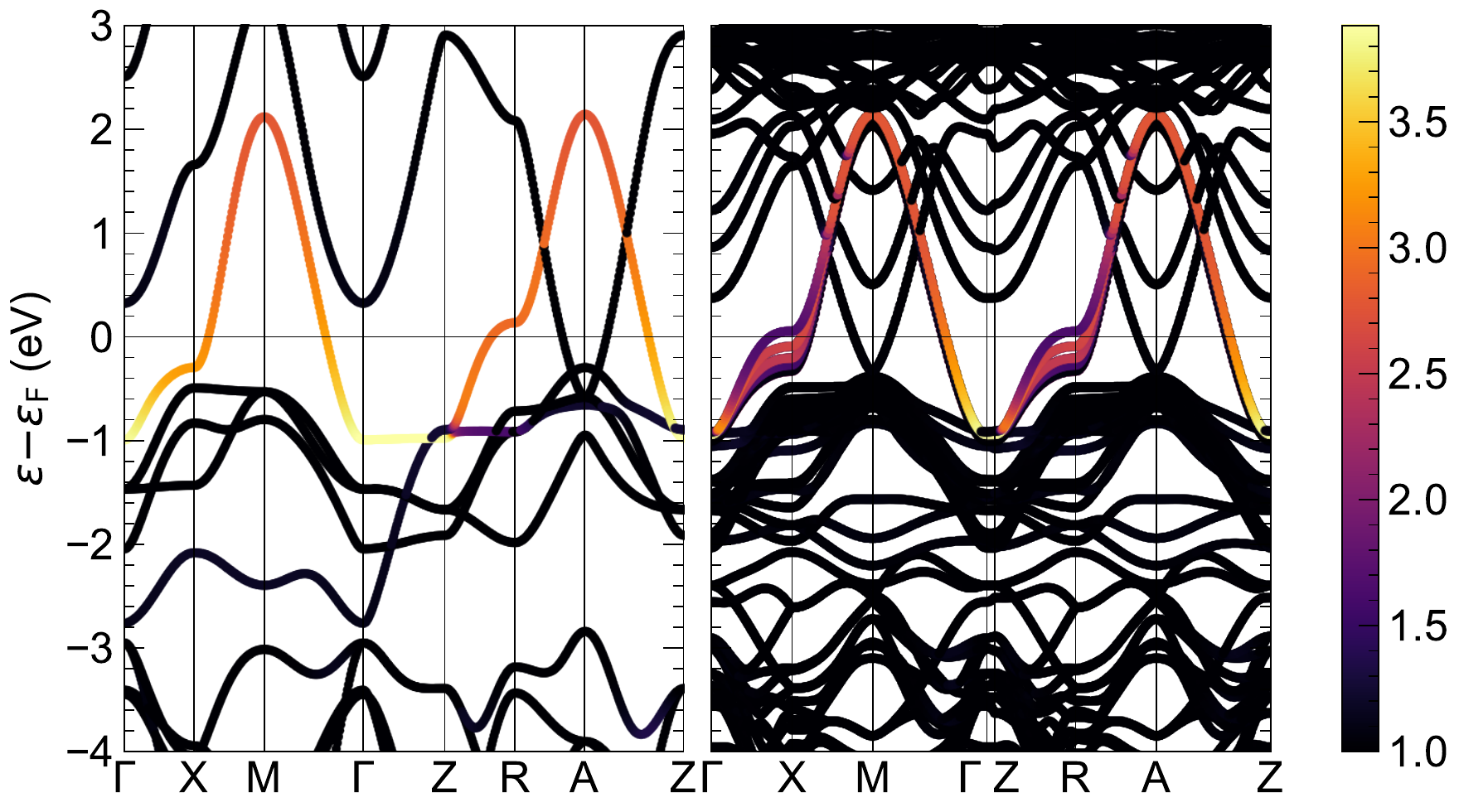}
    \caption{Mass enhancements derived from the electronic self-energy in the band basis ($\Sigma_{\nu\nu'}(\vb{k}, i\omega_{n})$) for 20\% hole-doped LaNiO$_2$ ($n=\infty$) (left) and La$_6$Ni$_5$O$_{12}$ ($n=5$) (right) (both at $d^{8.8}$ nominal filling).}
    \label{fig:band_mass}
\end{figure}

The orbital basis mass enhancements are derived from the diagonal parts of the electronic self-energy in the orbital basis $\Sigma_{mm'}(i\omega_{n})$ and describe the strength of correlations for a given orbital $m$. Of physical importance are the band basis mass enhancements, which describe the quasiparticle renormalization of the DFT bands and the amount of admixture of uncorrelated orbitals with the correlated orbitals. To obtain the band basis mass enhancements, we upfold the electronic self-energy from the orbital basis to the band basis via our projectors, 
\begin{equation}
    \Sigma_{\nu\nu'}(\vb{k}, i\omega_{n}) = \sum_{mm'} P_{\nu m}(\vb{k}) \Sigma_{mm'}(i\omega_{n}) P^{\dagger}_{\nu'm'}(\vb{k}),
\end{equation}
where $\Sigma_{\nu\nu'}(\vb{k}, i\omega_{n})$ is the self-energy in the band basis and $\nu$ are band indices. We then calculate the mass enhancements at every $\vb{k}$-point in the same fashion described in the main text. Figure \ref{fig:band_mass} shows the mass enhancements for each of the DFT bands, where the lighter color denotes a larger mass enhancement. We see that the Ni-$d_{x^{2}-y^{2}}$ band undergoes the largest renormalization as this is the most correlated orbital in both systems. The average mass enhancements for the $d_{x^{2}-y^{2}}$ band(s) are $\sim 2.8$ and $\sim 2.0-2.3$ for the 20\% hole-doped infinite-layer and 5-layer materials, respectively. The overall decrease of the band basis mass enhancements for both materials relative to the orbital basis mass enhancements indicates a significant admixture of the Ni-$d$ orbitals with the O-$p$ orbitals for both materials. The larger decrease in the 5-layer material indicates that there is slightly more admixture of the O-$p$ states than in the hole-doped infinite layer material.

\bibliography{ref.bib}
\end{document}